\documentclass[a4paper,11pt]{article}
\usepackage{jinstpub} % for details on the use of the package, please see the JINST-author-manual
\usepackage{lineno}
\usepackage{multirow}
\usepackage{siunitx}
\usepackage{comment}
\usepackage{multirow}
\usepackage{caption}
\usepackage{soul}

\title{\boldmath Updates on the DEAP-3600 experiment and steps towards the ARGO experiment}

\author{Susnata  Seth}
\affiliation{Department of Physics, Carleton University,\\
Ottawa, ON K1S 5B6, Canada}
\affiliation{Arthur B.  McDonald Canadian Astroparticle Physics Research Institute, Queen's University,\\
Kingston, ON K7L 2N6 , Canada}
\collaboration[c]{on behalf of GADMC and DEAP collaborations}

\emailAdd{SusnataSeth@cunet.carleton.ca}

\abstract{
The DEAP-3600 experiment, with an approximately 3.3-tonne liquid argon (LAr) target, is currently the world’s largest single-phase LAr dark matter detector. It is located 2 km underground at SNOLAB, Canada, one of the most radiopure underground laboratories.
With excellent pulse-shape discrimination against low-energy beta decays and precise position reconstruction, DEAP-3600 has set the most stringent WIMP–nucleon spin-independent cross-section exclusion limits for masses above 30 GeV/c$^{\rm 2}$  on argon and provided leading sensitivity to superheavy, multi-scattering dark matter candidate. Here we report the recent advances in understanding LAr properties and position reconstruction techniques using DEAP-3600 data along with hardware upgrades to mitigate residual challenging $\alpha$-backgrounds for WIMP search. As a part of Global Argon Dark Matter Collaboration (GADMC), the next-generation ARGO detector, featuring a 300-tonne fiducial LAr mass, is under development to significantly enhance sensitivity to rare dark matter interactions. Simulation-based studies of radiogenic neutron backgrounds and their mitigation strategies provide essential input to this design and will be described here.}

\begin{document}
\maketitle
\flushbottom

\section{Introduction}
\label{sec:intro}
Despite numerous compelling astrophysical and cosmological indications of dark matter (DM) inferred from its gravitational effects, uncovering its fundamental nature remains one of the major challenges in contemporary particle physics and astrophysics.  Substantial efforts are devoted to the direct detection of DM through interactions with detectors in terrestrial laboratories. Weakly Interacting Massive Particle (WIMP) remains a long-standing leading DM candidate, while Planck-mass DM candidate offers an equally intriguing and well-motivated alternative~\cite{BOZORGNIA2025671}.

DEAP-3600 is a single-phase LAr DM experiment located at SNOLAB, Canada. The experiment employs (3269 $\pm$ 24) kg  LAr target~\cite{DEAP:2023wri} contained within an acrylic vessel (AV) and viewed by 255 photomultiplier tubes (PMTs). The detector assembly is submerged in a water tank, primarily acts as an active muon veto and as shielding to neutron and gamma backgrounds from the cavern~\cite{AMAUDRUZ20191}.  The detector uses pulse-shape discrimination (PSD) to identify signal-like nuclear recoils (NRs)  and background-like electron recoils (ERs)~\cite{Adhikari_2021}.  To extend the sensitivity to DM-interactions,  the next-generation ARGO detector is under development. ARGO will contain a 400-tonne LAr target instrumented with digital silicon photomultipliers (SiPMs).  SNOLAB’s Cube Hall has been selected as the prospective site.  Here we summarize the recent results from the DEAP-3600 experiment and present studies aimed at optimizing ARGO detector design with a focus on neutron background mitigation.

\section{Recent results from DEAP-3600}
The DEAP-3600 detector was first filled with LAr in May 2016. Due to a leak in the neck of the detector, the LAr was contaminated with clean, radon-scrubbed nitrogen. The detector was subsequently emptied in August 2016 and refilled~\cite{AMAUDRUZ20191}. The DEAP-3600 experiment acquired data from its second fill between November 2016 and April 2020 and has recently resumed data taking following the completion of hardware upgrades.
With a total exposure of 758 tonne.days, collected  over 231 live-days between November 4, 2016 and October 31, 2017, DEAP-3600 has set a limit on WIMP-nucleon spin-independent cross-section on an argon target of 3.9 $\times$  10$^{-45}$ cm$^{2}$ for a 100 GeV/c$^{2}$ WIMP mass at 90\% C.L. ~\cite{PhysRevD.100.022004}. Additionally, due to its large size, DEAP-3600 is the first direct-detection experiment capable of probing Planck-mass scale DM candidate. The resulting constraint spans DM masses from 8.3 $\times$ 10$^{6}$ to  1.2 $\times$ 10$^{19}$ GeV/c$^{2}$ and excludes $^{40}$Ar-scattering cross-sections between 1.0 $\times$ 10 $^{-23}$ and 2.4  $\times$ 10 $^{-18}$ cm$^{2}$~\cite{PhysRevLett.128.011801}. 
The recent results from DEAP-3600 have expanded its physics reach through detailed studies of argon properties and by improving position reconstruction techniques required for  DM detection.  Notably, DEAP-3600 has measured the 
half-life of $^{39}$Ar to be ($302\; \pm \; 8_{\rm stat} \; \pm \;6_{\rm sys}$) years~\cite{DEAP:2025shk}.
An energy-dependent scintillation model for $\alpha$-particles has been developed  using  relative quenching  factor measurements in the (5-8) MeV region with DEAP-3600 data, together with a direct quenching measurement using a $^{210}$Po source by T. Doke et al.~\cite{DEAP:2024mov,DOKE1988291}. Both electronic and nuclear quenching factor have been extrapolated down to 10 keV and the associated systematic uncertainties have been evaluated individually. The resulting combined quenching factor is being applied in the currently ongoing WIMP search analysis using Profile Likelihood Ratio (PLR)  method and  the full second-fill dataset. In parallel, machine-learning based position reconstruction algorithm has been implemented, employing a feed-forward neural network with photoelectron pattern as input~\cite{Adhikari_2025}. This algorithm improves identification of the background events arising from the neck region of the detector.

\section{DEAP-3600 : Hardware upgrades and third fill}
The hardware upgrade has been completed to eliminate two challenging $\alpha$-background sources for the WIMP search, aiming to achieve a background-free sensitivity at 10$^{-46}$ cm$^{2}$ level for   a WIMP mass of 100 GeV/c$^{2}$ --- representing the first such achievement  for LAr based detector. The VUV scintillation light produced by $\alpha$-decays from $^{210}$Po isotope on the acrylic flowguides surface is absorbed by uncoated flowguides and also shadowed by the topology of the detector neck. Therefore, it produces background events within the WIMP region of interest (ROI), approximately (15.6 - 32.9) keV$_{\rm ee}$~\cite{PhysRevD.100.022004}.  Custom-developed pyrene-doped polystyrene coated new flowguides have been installed. The long decay time constant of pyrene and its wavelength shifting efficiency from UV photons to visible photons, enable this $\alpha$-background distinguishable from the NR band through PSD~\cite{GALLACHER2022166683,Garg:2023fzs}. Additionally, an external cooling system has been installed to supply LAr directly into the AV to prevent LAr film formation on the flowguides surfaces. Alpha-decays from  dust particulates dispersed in the LAr volume can produce dominant background events in the low energy region, due to degraded $\alpha$-energies and self-shadowing of scintillation light by the dust particulates. An extraction tube near the bottom of the AV has been installed to draw LAr, filter out the dust particulates and refill the AV with purified LAr. This filtration process will be performed over several cycles before final re-filling of the detector with LAr. Currently, the detector is refilled with LAr and  data taking is ongoing. The filtration of dust particulates is scheduled to begin soon.

\section{Future prospects and neutron backgrounds for ARGO}
The ARGO experiment is envisioned as the final milestone of the Global Argon Dark Matter Collaboration (GADMC). It is planned to use 400 tonnes of low‑radioactivity underground argon, of which 300 tonnes constitute the fiducial mass. With ten years of data collection, the experiment is expected to reach a WIMP search sensitivity approaching the so‑called ‘neutrino fog'~\cite{PhysRevLett.127.251802}.  Ongoing design and simulation studies are focused on assessment of radiogenic neutron backgrounds and developing strategies for their mitigation, which are critical for achieving the required background-free sensitivity, since  neutron interaction can mimic WIMP-induced signal. The requirement is fewer than one background event within the WIMP ROI and the fiducial mass over ten years of exposure.

Radiogenic neutrons with energies up to a few tens of MeV can originate from ($\alpha$,n) reactions. Alpha particles at MeV energy levels are produced from the decay chains of radioactive internal impurities of $^{238}$U, $^{235}$U, $^{232}$Th  in different detector components. The activity of these  decay chains for different detector components assumed in the ARGO simulation work are presented here, defining the target impurity levels for detector components  (shown in table~\ref{tab:activity_decaychain}). 
\begin{table}[tbp!]
    \centering
    \begin{tabular}{c|cccccc}
    \hline
    \multirow{2}{*}{Component} &
    \multicolumn{6}{|c}{Activity of Decay Chain~}\\
    \cline{2-7}
           & Unit & $^{\rm 238}$U &$^{\rm 238}$U&$^{\rm 238}$U& $^{\rm 232}$Th& $^{\rm 235}$U \\
          & &   upper & mid & lower & &\\
        \hline
         \hline
       protoDune-style cryostat & $\frac{\rm mBq}{\rm kg}$ & 4.2E+3 & 2.4E+3 & 1.7E+4 & 3.2E+3 & 1.9E+2\\
 Vacuum cryostat & $\frac{\rm mBq}{\rm kg}$ & 6.4E-1 & 2.0E-1 & 4.0E-1 & 1.1E+0 & 3.0E-2\\
    SiPM (only Silicon) & $\frac{\rm uBq}{\rm cm^{2}}$ & 1.6E-1 &1.6E-1 &3.2E+0 & 1.7E-1 &7.4E-3\\
       Acrylic vessel& ppb & 3.8E-3&  3.8E-3 & 0 & 6.7E-3  & 8.4E-4 \\
        \hline
    \end{tabular}
    \caption{Activities of $^{\rm 238}$U, $^{\rm 232}$Th and $^{\rm 235}$U  decay chains for different detector components.}
    \label{tab:activity_decaychain}
\end{table}
%%%%%%%%%%%%
\begin{figure}[tbp!]
\centering
\begin{minipage}[t]{.48\textwidth}
  \centering
  \includegraphics[width=\linewidth]{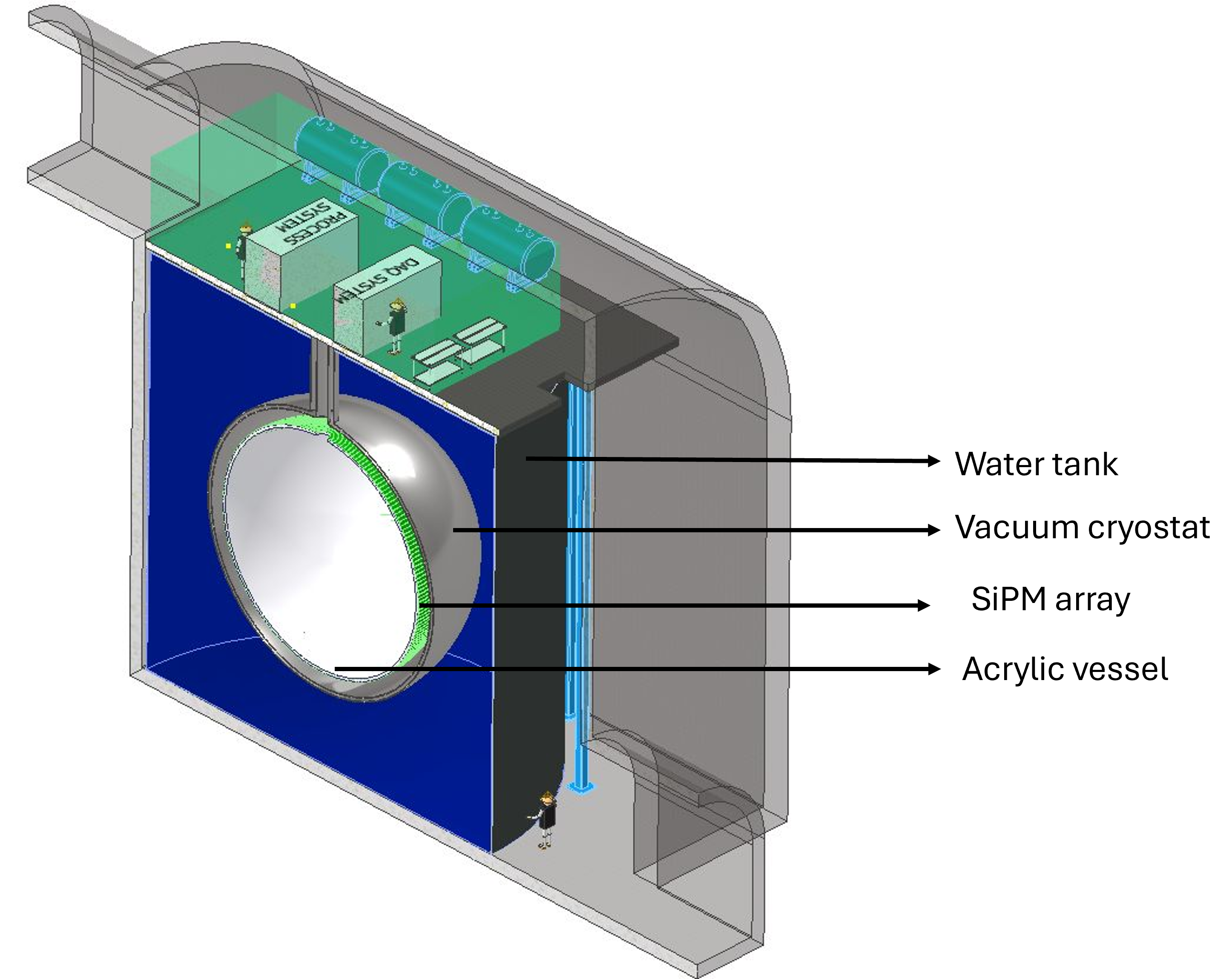}
  \captionof{figure}{ARGO Geometry B detector design showing key elements inside the Cube Hall at SNOLAB.}
  \label{fig:GeometrySphere}
\end{minipage}
  \hspace{1pc}%
\begin{minipage}[t]{.48\textwidth}
  \centering
   \includegraphics[width =\textwidth]{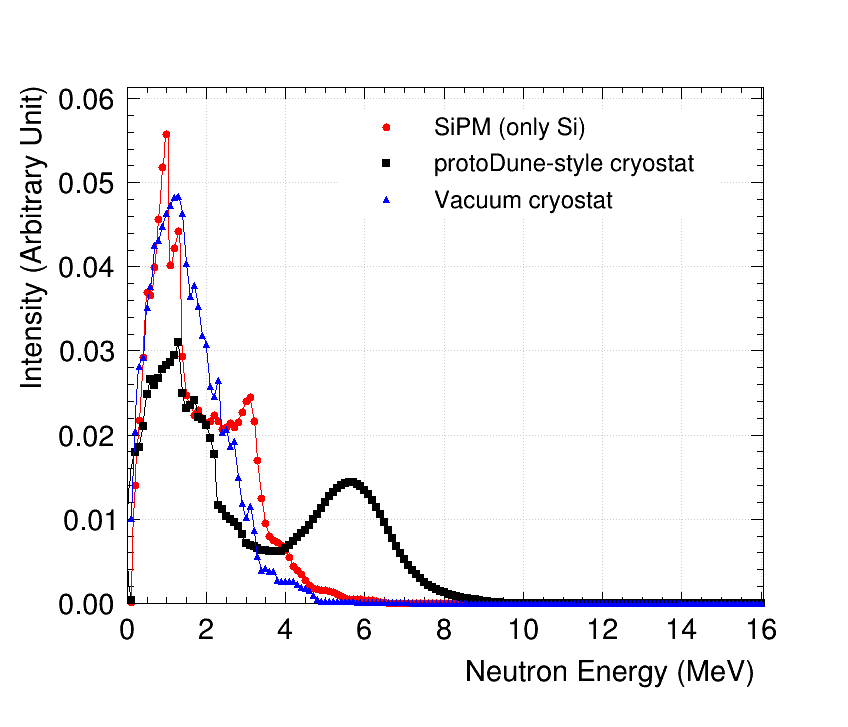}
  \captionof{figure}{Normalized neutron energy spectra from ($\alpha$,n)
reactions for different detector components.}
   \label{fig:n-spectra}
\end{minipage}
\end{figure}
%%%%%%%%%%%%%%%%%%%%%%%%%%%%%%%%%%%%%%%
Detailed Monte Carlo simulations were performed using RAT software framework based on GEANT4 and ROOT~\cite{rat}, to evaluate neutron leakage into the WIMP ROI over 3000 tonne$\cdot$year exposure. 
The conceptual design of the ARGO detector features  a spherical or cylindrical AV 
(both configurations have been considered yielding similar results on neutron background leakage) containing  400 tonnes of LAr, viewed by digital SiPMs covering the entire surface of AV and submerged in  LAr neutron veto. 
The detector is housed inside a cryostat surrounded by a water tank to reduce external neutron backgrounds from norite rock, following the same shielding strategy used in the DEAP-3600 experiment.  
Two geometries are considered. Geometry A contains a cylindrical AV with an inner diameter and height of about 7.1 m and a 15 cm wall thickness. It is immersed in a 1.6 m thick LAr veto inside a commercial protoDune-style cryostat,
consisting of two 1.3 cm thick stainless steel  shells separated by approximately 1.2 m of foam. Geometry B (see figure~\ref{fig:GeometrySphere}) employs a spherical AV with an inner diameter of approximately 8.1 m and a 10 cm wall thickness. It is submerged in  a 1 m thick LAr veto inside a custom-designed vacuum cryostat composed of two 3 cm thick stainless steel shells separated by 10 cm vacuum gap. In both cases, the inner AV surface is coated with a 3 $\mu$m   1,1,4,4-tetraphenyl-1,3-butadiene (TPB) wavelength shifter.  In both geometries the outer surface of the AV is covered in SiPM arrays. The stainless steel used in the vacuum cryostat is significantly more radiopure than that of the commercial protoDUNE-style cryostat.

Neutron energy spectra, and their spatial distributions are provided as inputs to the  RAT simulation, with neutrons emitted isotropically from their points of origin. The neutron yield   (neutron per decay per 100 keV) as a function of neutron energy from  ($\alpha$,n) reactions is calculated by NeuCBOT (Neutron Calculator Based On TALYS) tool~\cite{WESTERDALE201757}.  Given the  elemental composition and mass fraction of each material,  NeuCBOT  tool computes neutron yield  individually  for each material and  $^{238}$U, $^{235}$U, $^{232}$Th decay chain  using  ($\alpha$,n) total cross-section from  TALYS and mass stopping power from SRIM~\cite{refId0,SRIM-TRIM}.  It splits the $^{238}$U decay chain into three parts: all isotopes prior to $^{226}$Ra (half-life 1600 years), the segment beginning at $^{226}$Ra and continuing through its short-lived descendants, and a separate $^{210}$Pb sub-chain. Because $^{210}$Pb has a much longer half-life (22.2 years) than its precursors, it can accumulate through airborne $^{222}$Rn exposure, breaking secular equilibrium. The total energy spectra are built by weighting  neutron yields associated with each decay chain by their corresponding activities  and where appropriate, by the mass of the relevant detector components and  are then normalized to their total bin content (shown in figure~\ref{fig:n-spectra}). A conservative ($\alpha$,n)  neutron spectrum from rock, adopted from the DEAP-3600 design, is used for the ARGO design.

During  a neutron's passage  through the various detector  components, its energy is dissipated as NRs or ERs.  These  NRs originate from elastic scattering off  neutrons, while the ERs are produced  from  gamma-rays generated either through inelastic scattering or through capture of thermal neutrons at the end of the neutron track by various detector materials.  The total energy loss by  NRs and ERs have been recorded separately  for each physical volume defined in the simulation. The position for each neutron has been computed from energy-weighted argon-nucleus coordinates within the target, normalized by the total deposited energy. The optical simulation has  not been performed because it is too CPU intensive. The effect of PSD cuts has been estimated by tagging electron-recoil (ER) events with sufficient energy loss from ER energy in LAr target. Following selection criteria used to identify neutron leakage events are applied cumulatively. First, events with non-zero nuclear recoil (NR) energy deposition in the LAr target are  selected.  From these, NR events falling within the WIMP search ROI (15-35) keV$_{\rm ee}$, assuming a NR quenching factor of 0.25 in LAr, are retained. To suppress ERs, events with ER energy deposition in the LAr target exceeding 100 keV$_{\rm ee}$ are rejected. An additional ER event suppression is performed by rejecting events with total energy deposition in the LAr veto exceeding 100 keV$_{\rm ee}$. Fiducial constraints are then applied: the events occurring within 25 cm or 30 cm of the AV inner surface  are rejected for Geometry A and Geometry B, respectively. These fiducial cut preserve a   fiducial LAr mass of 300 tonne. Neutrons below 50 keV are not simulated for all cases, but a small number runs including thermal neutrons have been performed to  determine a correction factor for neutron capture gammas, applied at the last step of selection criteria when appropriate.

\begin{table}[tbh!]
  \centering
   % \centering
    \begin{tabular}{c|c|c}
    \hline
      \multirow{2}{*}{Neutron Source}&
      \multicolumn{2}{c}{Neutron Leakage in 3000 tonne$\cdot$year}\\
      \cline {2-3}
      &   Geometry A & Geometry B \\
      \hline
       Norite Rock  &  $<$ 0.13 (95\% C.L.) & $<$ 0.2 (95\% C.L.)\\
       Cryostat  & 42 $\pm$ 17.6 & 0.78 $\pm$ 0.09 \\
       SiPM (only Si) &  0.06 $\pm$ 0.03 & 0.105 $\pm$ 0.002 \\
       Acrylic vessel  &  0.72 $\pm$ 0.32 & 0.58 $\pm$ 0.02 \\
      \hline
        Total &   42.8 $\pm$ 17.6  & 1.5 $\pm$ 0.2\\
     \hline
    \end{tabular}
    \caption{Neutron leakage into WIMP ROI  in 3000 tonne$\cdot$year for ARGO detector.}
    \label{tab:n-leakage}
\end{table}

Optimization of water shield thickness, LAr veto thickness, AV thickness has been carried out to minimize  radiogenic neutron backgrounds within the WIMP ROI. 
Simulations varying the water shield thickness provides that a minimum 3 m and 2 m of water shield is required to prevent ($\alpha$,n) neutron leakage from rock in Geometry A and Geometry B, respectively. In Geometry B, incident neutrons are  uniformly generated within the stainless steel  cryostat shells, the SiPM and  the AV shell, whereas in Geometry A they are generated  from the innermost cryostat surface, the imaginary SiPM plane and uniformly  within the AV shell.  In both cases, rock neutrons have been generated from the rock and experimental hall boundaries. 
Table ~\ref{tab:n-leakage} shows that,  in the Cube Hall, the Geometry B detector design yields sufficiently low radiogenic neutron backgrounds over 3000 tonne$\cdot$year  exposure and  it benefits from the  low background custom-designed vacuum cryostat.

\section{Summary and outlook}
Data taking for the third and final fill of the DEAP-3600 experiment has begun. Fully automated analysis, including  PMT calibration, light yield determination and PSD for shadowed and dust $\alpha$-events,  is currently underway. The PLR analysis for WIMP search using the full second-fill data is 
now under internal review. In parallel, the design of ARGO detector, with the goal with achieving less than one neutron leakage for a 3000 tonne$\cdot$year exposure, is being investigated through  detailed Monte Carlo simulation and radiopure material selection. 
Additional studies on suppressing the neutron-induced events via identification of multi-scatter events, as well as, expanding the WIMP ROI into the low-energy region are also in progress.

%\section*{Acknowledgments}
\acknowledgments
We thank the Natural Sciences and Engineering Research Council of Canada (NSERC),
the Canada Foundation for Innovation (CFI),
the Ontario Ministry of Research and Innovation (MRI), 
and Alberta Advanced Education and Technology (ASRIP),
the University of Alberta,
Carleton University, 
Queen's University,
the Canada First Research Excellence Fund through the Arthur B.~McDonald Canadian Astroparticle Physics Research Institute,
Consejo Nacional de Ciencia y Tecnolog\'ia Project No. CONACYT CB-2017-2018/A1-S-8960, 
DGAPA UNAM Grants No. PAPIIT IN108020 and IN105923, 
and Fundaci\'on Marcos Moshinsky,
the European Research Council Project (ERC StG 279980),
the UK Science and Technology Facilities Council (STFC) (ST/K002570/1 and ST/R002908/1),
the Leverhulme Trust (ECF-20130496),
the Russian Science Foundation (Grant No. 21-72-10065),
the Spanish Ministry of Science and Innovation (PID2022-138357NB-C2) and the Community of Madrid (2018-T2/TIC-10494), 
the International Research Agenda Programme AstroCeNT (MAB/2018/7)
funded by the Foundation for Polish Science (FNP) from the European Regional Development Fund,
and the Polish National Science Centre (2022/47/B/ST2/02015).
Studentship support from
the Rutherford Appleton Laboratory Particle Physics Division,
STFC and SEPNet PhD is acknowledged.
We thank SNOLAB and its staff for support through underground space, logistical, and technical services.
SNOLAB operations are supported by the CFI
and Province of Ontario MRI,
with underground access provided by Vale at the Creighton mine site.
We thank Vale for their continuing support, including the work of shipping the acrylic vessel underground.
We gratefully acknowledge the support of the Digital Research Alliance of Canada,
Calcul Qu\'ebec,
the Centre for Advanced Computing at Queen's University,
and the Computational Centre for Particle and Astrophysics (C2PAP) at the Leibniz Supercomputer Centre (LRZ)
for providing the computing resources required to undertake this work.

% Bibliography

%% [A] Recommended: using JHEP.bst file
%% \bibliographystyle{JHEP}
%% \bibliography{biblio.bib}

%% or
%% [B] Manual formatting (see below)
%% (i) We suggest to always provide author, title and journal data or doi:
%% in short all the informations that clearly identify a document.
%% (ii) please avoid comments such as "For a review'', "For some examples",
%% "and references therein" or move them in the text. In general, please leave only references in the bibliography and move all
%% accessory text in footnotes.
%% (iii) Also, please have only one work for each \bibitem.

%\begin{thebibliography}{99}

%\bibitem{a}
%Author,
%\emph{Title},
%\emph{J. Abbrev.} {\bf vol} (year) pg.

%\bibitem{b}
%Author,
%\emph{Title},
%arxiv:1234.5678.

%\bibitem{c}
%Author,
%\emph{Title},
%Publisher (year).

%\end{thebibliography}
\bibliography{biblio}
\end{document}